\documentclass[12pt]{article}
\usepackage{graphicx}
\usepackage[cp1251]{inputenc}
\usepackage{rotating}
\begin{document}
 \noindent {\footnotesize\it Astronomy Reports, 2017, Vol. 61, No 9, pp. 727--738}
 \newcommand{\dif}{\textrm{d}}

 \noindent
 \begin{tabular}{llllllllllllllllllllllllllllllllllllllllllllll}
 & & & & & & & & & & & & & & & & & & & & & & & & & & & & & & & & & & & & & \\\hline\hline
 \end{tabular}

  \vskip 0.5cm
 \centerline{\bf\large Galactic Orbits of Selected Companions of the Milky Way}
 \bigskip
 \bigskip
  \centerline
 {
 A.T. Bajkova,
 V.V. Bobylev
 }
 \bigskip
{\small \it
Central (Pulkovo) Astronomical Observatory, Russian
Academy of Sciences, Pulkovskoe shosse 65, St. Petersburg, 196140
Russia
 }
 \bigskip
 \bigskip
 \bigskip

 {
{\bf Abstract}---High-accuracy absolute proper motions, radial
velocities, and distances have now been measured for a number of
dwarf-galaxy companions of the Milky Way, making it possible to
study their 3D dynamics. Galactic orbits for 11 such galaxies
(Fornax, Sagittarius, Ursa Minor, LMC, SMC, Sculptor, Sextans,
Carina, Draco, Leo I, Leo II) have been derived using two
previously refined models for the Galactic potential with the
Navarro–Frenk–White and Allen–Santill\'an expressions for the
potential of the dark-matter halo, and two different masses for
the Galaxy within 200 kpc—$0.75\times 10^{12}M_\odot$ and
$1.45\times 10^{12}M_\odot$. The character of the orbits of most
of these galaxies indicates that they are tightly gravitationally
bound to the Milky Way, even with the lower-mass model for the
gravitational potential. One exception is the most distant galaxy
in the list, Leo I, whose orbit demonstrates that it is only
weakly gravitationally bound, even using the higher-mass model of
the gravitational potential.
  }

\medskip DOI: 10.1134/S1063772917080017

 \section{INTRODUCTION}
With a size of about 1000 kpc, the Local Group contains more than
60 galaxies [1], with the Andromeda Galaxy and the Milky Way
dominating. Each of these two galaxies is surrounded by a cloud of
companion galaxies. The following estimated masses for the Local
Group (LG), Andromeda Galaxy (M31), and Milky Way (MW) were
derived in [2] from data on the Local Group galaxies, assuming
that the group is in equilibrium:
$M_{LG}=(2.5\pm0.4)\times10^{12}M_\odot$,
 $M_{M31}=(1.7\pm0.3)\times10^{12}M_\odot$ and
  $M_{MW}=(0.8\pm0.5)\times10^{12}M_\odot$.

The volume bounded by the virial radius of the Milky Way,
$r_{vir}\sim300$~kpc, contains more than 40 known dwarf galaxies
[3]. Analysis of the distances and radial velocities of dwarf
galaxies in this volume indicates that they are gravitationally
bound to the Milky Way. One exception is one of the most distant
galaxies of the group, Leo I [1,4,5]. Using the total space
velocities of the galaxies can clarify the question of whether
Local Group members are gravitationally bound, and hence of the
mass of the Milky Way. However, unlike uncertainties in radial
velocities, proper motion uncertainties depend on the distance, so
that angular measurements with very high accuracy are needed.

Nevertheless, analysis of the total space velocities of several
dwarf galaxies that are companions to the Milky Way have been
repeatedly performed by different sets of authors with different
aims [6–8]. For example, it was found that the estimate of the
total mass of the Milky Way depends considerably on the dwarf
galaxy Leo I [4]; the question of whether the dwarf galaxy Leo II
is gravitationally bound to the Milky Way was considered in [9];
the frequency of mutual collisions of the companion galaxies was
estimated in [6]; and the most probable mass of the Galaxy within
a sphere 200 kpc in radius was estimated to be
$M_{200}=1.1\times10^{12}M_\odot$~[8].

The absolute proper motions of several dwarf companions of the
Milky Way were measured using the Hubble Space Telescope (HST).
These data, together with ground-based observations, are collected
in [10]. Analyzing these data is of considerable interest for
studying the dynamics of the Local Group of galaxies [11].

In [12], we re-determined parameters of the three most popular
three-component (bulge, halo, disk) axially symmetric models for
the Galaxy’s gravitational potential, which differ in the type of
expression used for the dark-matter halo, namely, the expressions
of Allen and Santill\'an (I) [13], Wilkinson and Evans (II) [14],
and Navarro, Frenk, and White (III) [15]. In all these models, the
bulge and disk are described using expressions from Miyamoto and
Nagai [16]. We used various modern observations over a wide range
of Galactocentric distances R between 0 and 200 kpc. For distances
of about 20 kpc, we used radial velocities of hydrogen clouds at
tangential points and data on 130 masers with measured
trigonometric parallaxes; at larger distances, we used averaged
rotation velocities for a variety of objects (carbon stars,
giants, globular clusters, and dwarf galaxies) from the survey
[17]. We fitted the observed velocities with model rotation curves
taking into account limits for the local density of matter and the
vertical force. It is important to note that the models we
obtained correspond to different total masses for the Galaxy. The
Galaxy's mass within 200 kpc is highest for Model I,
 $M_G(R\leq200$~kpc$)=(1.45\pm0.30)\times10^{12}M_\odot$, and lowest for Model II,
 $M_G(R\leq200$~kpc$)=(0.61\pm0.12)\times10^{12}M_\odot.$ Model III
can be considered the best of the models considered, as it
provides the lowest residuals between the data and the model
rotation curve. The Galaxy's mass in Model III
 $M_G(R\leq200$~kpc$)=(0.75\pm0.19)\times10^{12}M_\odot.$

Since the positions and velocities of the listed dwarf companion
galaxies are continuously being refined, it is of considerable
interest to analyze their 3D dynamics using the refined Galactic
potential in order to establish the extent to which the companion
galaxies are bound to our Galaxy. This is the subject of the
present study.

 \section{DATA}
The most important source of data for our study is [10], in which
mean proper motions are derived for 11 galactic companions of the
Milky Way. For each of these galaxies, from one to four
independent measurements of absolute proper motions were used, at
least one derived from HST observations. The galaxies’
coordinates, distances, and radial velocities from [1] were used
in [10].

Our list contains 11 dwarf-galaxy companions of the Milky Way:
Fornax, Sagittarius, Ursa Minor, LMC, SMC, Sculptor, Sextans,
Carina, Draco, Leo I, Leo II. The data for seven of these galaxies
were taken directly from [10] and [1].

We derived new mean absolute proper motions for Draco, using two
independent determinations of the absolute proper motions for this
galaxy. The first resulted from the reduction of HST observations:
 $\mu_\alpha\cos \delta= 0.177\pm0.063$ milliarcsec/year (mas/year) and
  $\mu_\delta=-0.221\pm0.063$ mas/year [18]. The second measurement is based on
ground-based observations with the Subaru telescope:
 $\mu_\alpha\cos \delta=-0.284\pm0.047$ mas/year and
 $\mu_\delta=-0.289\pm0.014$ mas/year [19]. The mean of these two measurements is
  $\mu_\alpha\cos \delta=-0.054\pm0.055$ mas/year and
   $\mu_\delta=-0.255\pm0.042$ mas/year. We
adopted the heliocentric distance $d=82.4\pm5.8$ kpc from the
estimate [20], based on an analysis of 270 RR Lyrae variables, and
the heliocentric radial velocity $V_r=-293.3\pm1.0$ km/s [21].

We used the new mean proper motions for the Magellanic Clouds (LMC
and SMC) from [22]. These are based on observations with the HST
[23] and an analysis of stellar proper motions from the Gaia DR1
catalog [24, 25].

Two sets of measurements are available for the dwarf galaxy Leo
II, which we decided not to average. The absolute proper motions
of this galaxy were obtained in [9] using HST data with an epoch
difference of about 14 years: $\mu_\alpha\cos
\delta=0.104\pm0.113$ mas/year and $\mu_\delta=-0.033\pm0.151$
mas/year. When computing these values, measurements of 3224 stars
and 17 galaxies were used to derive the absolute motions. We took
the coordinates, radial velocity $V_r=78.0\pm0.1$ km/s, and
distance $d=233\pm14$ kpc from[1]. We will refer to this data set
as Leo II (a).

The absolute proper motions of Leo II were also measured in [26]
from HST data with an epoch difference of about 2 years:
$\mu_\alpha\cos \delta=-0.069\pm0.037$ mas/year and
$\mu_\delta=-0.087\pm0.039$ mas/year. The reduction to absolute
proper motions was based on measurements of more than 100
background galaxies and two quasars. Piatek et al. [26] used the
radial velocity $V_r=79.1\pm0.6$ km/s [27] and distance
$d=233\pm15$ kpc [28] for Leo II. They suggested that they had
achieved higher accuracy in their study by averaging six
independent measurements. We will refer to this data set as Leo II
(b). These two data sets have considerably different proper
motions and random errors, leading to our decision to consider
both.

The basic parameters of the 11 dwarf galaxies are collected in
Table 1, whose columns contain each galaxy’s (1) name, (2)--(3)
Galactic coordinates $l$ and $b$, (4)--(5) proper motions
$\mu_\alpha\cos\delta$ and $\mu_\delta,$ (6) heliocentric radial
velocity $V_r,$ and (7) heliocentric distance $d$.

 {\begin{table}[t]                                    
 \caption[]
 {\small\baselineskip=1.0ex
 Data used for companion galaxies of the Milky Way
  }
 \label{t:1}
 \small\baselineskip=1.0ex
\begin{center}\begin{tabular}{|l|r|r|r|r|r|r|r|}\hline
 Galaxy     & $l,$ & $b,$ & $\mu_\alpha\cos\delta,$ & $\mu_\delta,$ & $V_r,$ & $d,$ \\
            & deg  & deg  & mas/year & mas/year & km/s & kpc \\\hline
 Sgr        & $  5.57$ & $-14.17$ & $-2.684\pm0.200$ & $-1.015\pm0.070$ & $ 140.0\pm2.0$ & $ 26\pm2$ \\
 LMC        & $280.47$ & $-32.89$ & $ 1.903\pm0.014$ & $ 0.235\pm0.026$ & $ 262.2\pm3.4$ & $ 50\pm3$ \\
 SMC        & $302.81$ & $-44.33$ & $ 0.723\pm0.025$ & $-1.172\pm0.044$ & $ 145.6\pm0.6$ & $ 64\pm4$ \\
 Draco      & $ 86.37$ & $+34.71$ & $-0.054\pm0.055$ & $-0.255\pm0.042$ & $-293.3\pm1.0$ & $ 82\pm6$ \\
 UMi        & $104.95$ & $+44.80$ & $-0.036\pm0.071$ & $ 0.144\pm0.083$ & $-246.9\pm0.1$ & $ 76\pm3$ \\
 Sculptor   & $287.53$ & $-83.16$ & $ 0.168\pm0.104$ & $-0.036\pm0.112$ & $ 111.4\pm0.1$ & $ 86\pm6$ \\
 Sextans    & $243.50$ & $+42.27$ & $-0.260\pm0.410$ & $ 0.100\pm0.440$ & $ 224.2\pm0.1$ & $ 86\pm4$ \\
 Carina     & $260.11$ & $-22.22$ & $ 0.220\pm0.090$ & $ 0.150\pm0.090$ & $ 222.9\pm0.1$ & $105\pm6$ \\
 Fornax     & $237.25$ & $-65.66$ & $ 0.493\pm0.041$ & $-0.351\pm0.037$ & $  53.3\pm0.8$ & $139\pm8$ \\
 Leo II (a) & $220.17$ & $+67.23$ & $ 0.104\pm0.113$ & $-0.033\pm0.151$ & $  78.0\pm0.1$ & $233\pm14$\\
 Leo II (b) & $220.17$ & $+67.23$ & $-0.069\pm0.037$ & $-0.087\pm0.039$ & $  79.0\pm0.6$ & $233\pm15$\\
 Leo I      & $225.99$ & $+49.11$ & $-0.114\pm0.030$ & $-0.126\pm0.029$ & $ 282.9\pm0.5$ & $257\pm13$\\
 \hline
 \end{tabular}\end{center}\end{table}}

 {\begin{table}[t]                                    
 \caption[]
 {\small\baselineskip=1.0ex
Parameters of Models I and III for the Galactic potential
according to [12], for $M_{gal}=2.325\times 10^7 M_\odot$
  }
 \label{t:model-III}
 \begin{center}\begin{tabular}{|c|r|r|}\hline
 Parameter        & Model I & Model III\\\hline
 $M_b$(M$_{gal}$) & 386$\pm10$     &    443$\pm27$  \\
 $M_d$(M$_{gal}$) & 3092$\pm62$    &   2798$\pm84$ \\
 $M_h$(M$_{gal}$) & 452$\pm83$    &  12474$\pm3289$ \\
 $b_b$(kpc)       & 0.2487$\pm0.0060$ & 0.2672$\pm0.0090$ \\
 $a_d$(kpc)       & 3.67$\pm0.16$   &   4.40$\pm0.73$ \\
 $b_d$(kpc)       & 0.3049$\pm0.0028$ & 0.3084$\pm0.0050$ \\
 $a_h$(kpc)       & 1.52$\pm0.18$    &    7.7$\pm2.1$ \\\hline
 \end{tabular}\end{center}\end{table}}
 {\begin{table}[t]                                    
 \caption[]
 {\small\baselineskip=1.0ex
  Initial velocities (in km/s) in the fixed Cartesian $U,V,W$ and cylindrical $\Pi,\Theta$ coordinates
  }
 \label{t:UVW}
 \small\baselineskip=1.0 ex
\begin{center}\begin{tabular}{|l|c|c|c|c|c|c|c|}\hline
 Galaxy     & $U$ & $V$ & $W$ & $\Pi$ & $\Theta$ \\\hline
 Sgr        & $ 232\pm2~$ & $  30\pm25$ & $ 219\pm26$ & $ 234\pm4 $ & $   3\pm25$ \\
 LMC        & $ -42\pm11$ & $-217\pm11$ & $ 229\pm16$ & $ 217\pm11$ & $  38\pm11$ \\
 SMC        & $  20\pm23$ & $-158\pm23$ & $ 161\pm18$ & $ 154\pm23$ & $  44\pm23$ \\
 Draco      & $  93\pm17$ & $  -6\pm15$ & $-137\pm19$ & $ -12\pm15$ & $  92\pm17$ \\
 UMi        & $   7\pm4~$ & $  94\pm26$ & $-186\pm28$ & $  83\pm24$ & $  43\pm11$ \\
 Sculptor   & $- 34\pm61$ & $ 193\pm46$ & $- 99\pm4~$ & $-155\pm50$ & $ 121\pm58$ \\
 Sextans    & $-168\pm100$& $118\pm160$ & $ 117\pm180$& $ -9\pm145$ & $ 206\pm121$ \\
 Carina     & $- 73\pm27$ & $  13\pm15$ & $  39\pm52$ & $   6\pm16$ & $  74\pm26$ \\
 Fornax     & $- 35\pm11$ & $-130\pm7~$ & $ 107\pm12$ & $  123\pm8$ & $ -55\pm9~$ \\
 Leo II (a) & $112\pm197$ & $254\pm199$ & $ 109\pm51$ & $-253\pm198$& $ 115\pm198$\\
 Leo II (b) & $ -40\pm56$ & $ 124\pm54$ & $  41\pm21$ & $ -43\pm55$ & $ 123\pm54$\\
 Leo I      & $-168\pm37$ & $ -32\pm38$ & $  94\pm33$ & $ 143\pm38$ & $  93\pm38$\\
  \hline
 \end{tabular}\end{center}\end{table}}
 {\begin{table}[t]                                    
 \caption[]
 {\small\baselineskip=1.0ex
Parameters of the orbits of the dwarf-galaxy companions of the
Milky Way, shown in Figs. 1--4 for Model III (top) and Model I
(bottom).
 }
 \label{t:ecc}
 \small\baselineskip=1.0ex
\begin{center}\begin{tabular}{|l|r|r|r|}\hline
 Galaxy  & $a_{min}$ & $a_{max}$ & $e$ \\
            & kpc & kpc &    \\\hline
 Sgr        & $ 15  $ & $ 56   $ & $0.58$ \\
 LMC        & $ 50  $ & $ 912  $ & $0.90$ \\
 SMC        & $ 61  $ & $ 136  $ & $0.38$ \\
 Draco      & $ 46  $ & $ 119  $ & $0.45$ \\
 UMi        & $ 68  $ & $ 170  $ & $0.43$ \\
 Sculptor   & $ 75  $ & $ 268  $ & $0.57$ \\
 Sextans    & $ 82  $ & $ 460  $ & $0.70$ \\
 Carina     & $ 29  $ & $ 107  $ & $0.57$ \\
 Fornax     & $ 129 $ & $ 286  $ & $0.38$ \\
 Leo II (a) & $ 234 $ & $ 2457 $ & $0.83$ \\
 Leo II (b) & $ 225 $ & $  362 $ & $0.23$ \\
 Leo I      & $ 98  $ & $ 1245 $ & $0.85$ \\ \hline
 LMC        & $ 48  $ & $  225 $ & $0.65$ \\
 Leo I      & $ 79  $ & $  975 $ & $0.85$ \\
 \hline
 \end{tabular}\end{center}\end{table}}
 {\begin{table}[t]                                    
 \caption[]
 {\small\baselineskip=1.0ex
Orbital characteristics of the dwarf-galaxy companions of the
Milky Way, computed taking into account the uncertainties in the
input data and the parameters of the gravitational potential for
Model III (top) and Model I (bottom).
 %
 }
 \label{t:emc}
 \small\baselineskip=1.0ex
\begin{center}\begin{tabular}{|l|r|r|r|}\hline
 Galaxy  & $a_{min}$ & $a_{max}$ & $e$ \\
            & kpc & kpc &    \\\hline
 Sgr        & $ 15\pm2  $ & $ 62\pm22   $ & $0.59\pm0.07$ \\
 LMC        & $ 48\pm2  $ & $ 920\pm408  $ & $0.88\pm0.07$ \\
 SMC        & $ 60\pm3  $ & $ 160\pm75  $ & $0.41\pm$0.15 \\
 Draco      & $ 45\pm8  $ & $ 122\pm23  $ & $0.45\pm0.06$ \\
 UMi        & $ 68\pm7  $ & $ 218\pm100  $ & $0.48\pm0.13$ \\
 Sculptor   & $ 71\pm13  $ & $ 450\pm432  $ & $0.61\pm0.16$ \\
 Sextans    & $ 73\pm21  $ & $ 1933\pm1374  $ & $0.83\pm0.08$ \\
 Carina     & $ 38\pm28  $ & $ 112\pm13  $ & $0.54\pm$0.23 \\
 Fornax     & $ 121\pm21 $ & $ 363\pm209  $ & $0.44\pm$0.14 \\
 Leo II (a) & $ 221\pm32 $ & $ 2521\pm1185 $ & $0.81\pm0.09$ \\
 Leo II (b) & $ 192\pm60 $ & $  573\pm378 $ & $0.44\pm0.18$ \\
 Leo I      & $ 103\pm35  $ & $ 1323\pm289 $ & $0.85\pm0.05$ \\ \hline
 LMC        & $ 48\pm2  $ & $  239\pm79 $ & $0.65\pm0.08$ \\
 Leo I      & $ 78\pm30  $ & $  997\pm348 $ & $0.84\pm0.07$ \\
 \hline
 \end{tabular}\end{center}\end{table}}

 \section{METHOD}
 \subsection{Model for the Galactic Potential}
The axially symmetric gravitational potential of the Galaxy was
represented as the sum of three components --- the central,
spherical bulge $\Phi_b(r(R,Z))$, the disk $\Phi_d(r(R,Z))$, and
the massive, spherical dark-matter halo $\Phi_h(r(R,Z))$:
 \begin{equation}
 \begin{array}{lll}
  \Phi(R,Z)=\Phi_b(r(R,Z))+\Phi_d(r(R,Z))+\Phi_h(r(R,Z)).
 \label{pot}
 \end{array}
 \end{equation}
Here, we used a cylindrical coordinate system ($R,\psi,Z$) with
its origin at the Galactic center. In Cartesian coordinates
$(X,Y,Z)$ with their origin at the Galactic center, the distance
to a star (the spherical radius) is $r^2=X^2+Y^2+Z^2=R^2+Z^2,$
where the $X$ axis is directed from the Sun toward the Galactic
center, the Y axis is perpendicular to the $X$ axis and points in
the direction of the Galactic rotation, and the $Z$ axis is
perpendicular to the Galactic $(XY)$ plane and points in the
direction of the North Galactic pole. The gravitational potential
is expressed in units of 100 km$^2$ s$^{-2}$, distances in kpc,
masses in units of the mass of the Galaxy, $M_{gal}=2.325\times
10^7 M_\odot$, and the gravitational constant is taken to be
$G=1.$

We expressed the potentials of the bulge, $\Phi_b(r(R,Z))$, and
disk, $\Phi_d(r(R,Z))$, in the form suggested by Miyamoto and
Nagai [16]:
 \begin{equation}
  \Phi_b(r)=-\frac{M_b}{(r^2+b_b^2)^{1/2}},
  \label{bulge}
 \end{equation}
 \begin{equation}
 \Phi_d(R,Z)=-\frac{M_d}{\Biggl[R^2+\Bigl(a_d+\sqrt{Z^2+b_d^2}\Bigr)^2\Biggr]^{1/2}},
 \label{disk}
\end{equation}
where $M_b, M_d$ are the masses of these components, and $b_b,
a_d, b_d$ are the scale parameters of the components in kpc. For
the halo component, we used the expression presented in [15]:
 \begin{equation}
  \Phi_h(r)=-\frac{M_h}{r} \ln {\Biggl(1+\frac{r}{a_h}\Biggr)}.
 \label{halo-III}
 \end{equation}
The right column of Table 2 presents the model parameters of the
Galactic potential (2)--(4) computed in [12] using the rotation
velocities of Galactic objects at distances $R$ within $\sim200$
kpc. Note that the corresponding Galactic rotation curve was
derived using the local parameters $R_\odot=8.3$ kpc and
$V_\odot=244$ km/s.

The model (2)--(4) is called Model III in [12]. In our present
study, Model III is our main model, and is regarded to be optimal
from the point of view of its fit to the observations. Because of
this, we computed almost all the dwarf-galaxy orbits using this
particular model. In the model, the total mass of the Galaxy
within a sphere of radius 200 kpc is
$M_{200}=(0.75\pm0.19)\times10^{12}M_\odot.$

Model II of Wilkinson and Evans, also considered in [12], is close
to Model III. The total mass of the Galaxy in this model is
comparable to the mass for Model III,
$M_{200}=(0.61\pm0.12)\times10^{12}M_\odot.$ The similarity of
Models II and III can be demonstrated by integrating the orbits of
selected globular clusters.

For this reason, Model II is not very important for us, and we do
not consider it farther. Model I [12] is far more interesting for
us, as it features a total mass of the Galaxy within 200 kpc that
is almost twice as large:
$M_{200}=(1.45\pm0.30)\times10^{12}M_\odot$. In this model, the
potential of the dark matter halo is expressed [29] as:
\begin{equation}
 \renewcommand{\arraystretch}{3.2}
  \Phi_h(r) = \left\{
  \begin{array}{ll}\displaystyle
  \frac{M_h}{a_h}\biggl( \frac{1}{(\gamma-1)}\ln \biggl(\frac{1+(r/a_h)^{\gamma-1}}{1+(\Lambda/a_h)^{\gamma-1}}\biggr)-
  \frac{(\Lambda/a_h)^{\gamma-1}}{1+(\Lambda/a_h)^{\gamma-1}}\biggr),
  &\textrm{if } r\leq \Lambda, \\\displaystyle
  -\frac{M_h}{r} \frac{(\Lambda/a_h)^{\gamma}}{1+(\Lambda/a_h)^{\gamma-1}}, &\textrm{if }  r>\Lambda,
  \end{array} \right.
 \label{halo-I}
 \end{equation}
where $M_h$ is the mass and ah a scaling parameter, the distance
from the Galactic center is taken to be $\Lambda=200$ kpc, and the
dimensionless coefficient is $\gamma=2.0$. The parameters of this
model are also given in Table 2. With Eqs. (2)--(3) taken into
account, we have a modification of the model of Allen and
Santill\'an [13]. We applied this model to analyze the orbits of
the Large Magellanic Cloud (LMC) and the most distant dwarf galaxy
in our list, Leo I.

 \subsection{Construction of the Orbits}
The equation of motion of a test particle in an axially symmetric
gravitational potential can be obtained from the Lagrangian of the
system $\pounds$ (see Appendix A in [29]):
\begin{equation}
 \begin{array}{lll}
 \pounds(R,Z,\dot{R},\dot{\psi},\dot{Z})=\\
 \qquad0.5(\dot{R}^2+(R\dot{\psi})^2+\dot{Z}^2)-\Phi(R,Z).
 \label{Lagr}
 \end{array}
\end{equation}
Introducing the canonical moments
\begin{equation}
 \begin{array}{lll}
    p_{R}=\partial\pounds/\partial\dot{R}=\dot{R},\\
 p_{\psi}=\partial\pounds/\partial\dot{\phi}=R^2\dot{\psi},\\
    p_{Z}=\partial\pounds/\partial\dot{Z}=\dot{Z},
 \label{moments}
 \end{array}
\end{equation}
we obtain the Lagrangian equations in the form of a system of six
first-order differential equations:
 \begin{equation}
 \begin{array}{llllll}
 \dot{R}=p_R,\\
 \dot{\psi}=p_{\psi}/R^2,\\
 \dot{Z}=p_Z,\\
 \dot{p_R}=-\partial\Phi(R,Z)/\partial R +p_{\psi}^2/R^3,\\
 \dot{p_{\psi}}=0,\\
 \dot{p_Z}=-\partial\Phi(R,Z)/\partial Z.
 \label{eq-motion}
 \end{array}
\end{equation}
We integrated Eqs. (8) using a fourth-order Runge-Kutta algorithm.

The Sun’s peculiar velocity with respect to the Local Standard of
Rest was taken to be
$(u_\odot,v_\odot,w_\odot)=(11.1,12.2,7.3)\pm(0.7,0.5,0.4)$ km/s
[30]. Here, we used the heliocentric velocities in a moving
Cartesian coordinate system with $u$ directed towards the Galactic
center, $v$ in the direction of the Galactic rotation, and $w$
perpendicular to the Galactic plane and directed towards the north
Galactic pole.

Let us denote the initial positions and proper motions of a test
particle in the Galactocentric coordinate system as
$(x_o,y_o,z_o,u_o,v_o,w_o)$. The initial positions and velocities
of a test particle in the Cartesian Galactic coordinate system are
then given by the formulas:
\begin{equation}
 \begin{array}{llllll}
 X=R_0-x_o, Y=y_o, Z=z_o,\\
 U=u_o+u_\odot,\\
 V=v_o+v_\odot+V_0,\\
 W=w_o+w_\odot,\\
 \Pi=-U\cos\psi_o+V\sin\psi_o,\\
 \Theta=U\sin\psi_o+V\cos\psi_o,
 \label{init}
 \end{array}
\end{equation}
where $R_0$ and $V_0$ are the Galactocentric distance and the
linear velocity of the Local Standard of Rest around the Galactic
center, and tan $\tan\psi_o=Y/X$.

The initial space velocities $U,V,W$ and the velocities $\Pi$ and
$\Theta$ along the $R$ and $\psi$ cylindrical coordinates are
collected in Table 3. Table 4 presents the perigalactic,
$a_{min}$, and apogalactic, $a_{max}$, distances and
eccentricities $e$ of the derived orbits of the dwarf companion
galaxies.

\begin{figure}[p]
{\begin{center}
 \includegraphics[width=0.9\textwidth]{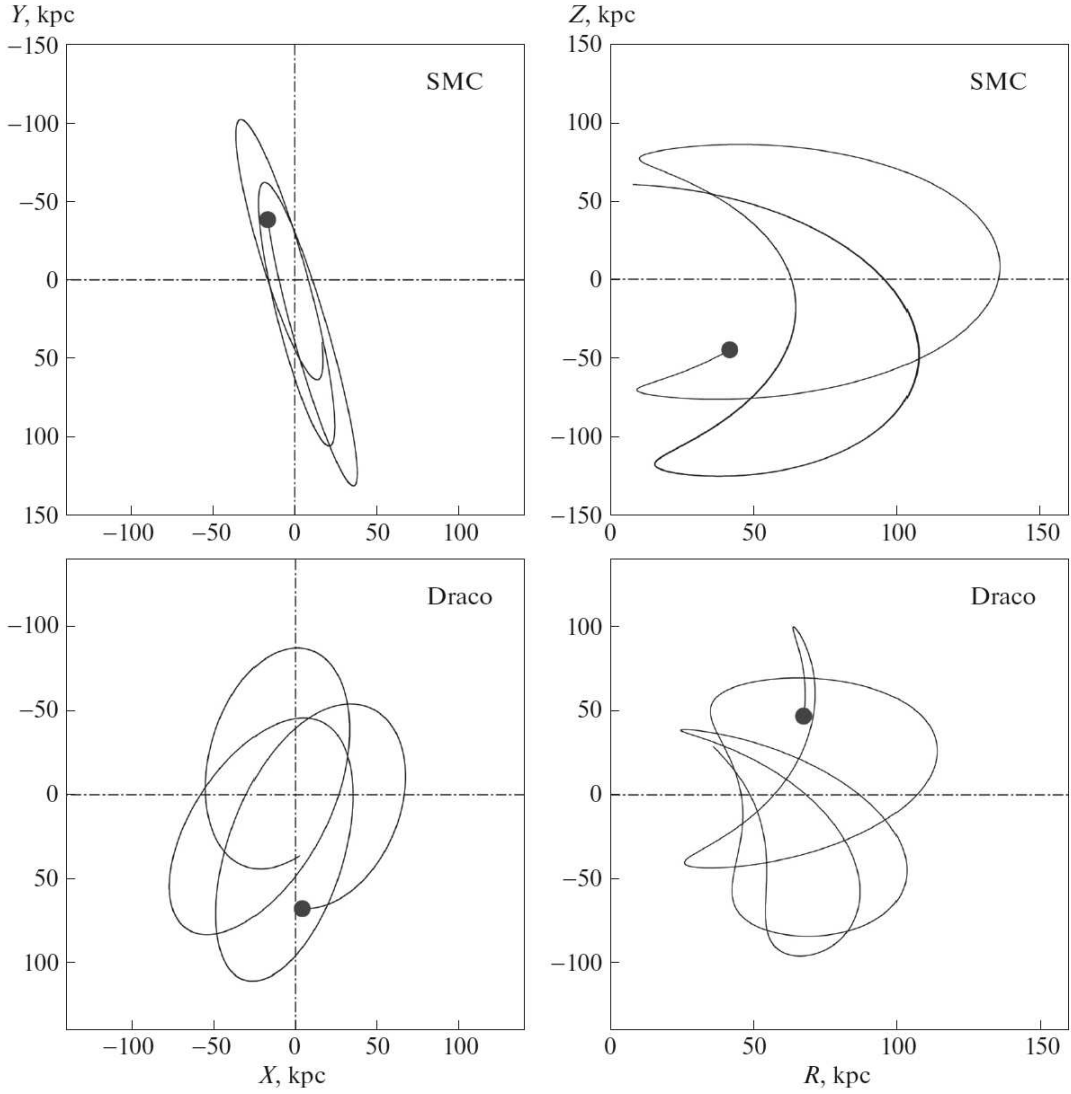}
 \caption{
Galactic orbits of the dwarf galaxies SMC and Draco over 10
billion years into the past.
  } \label{f123}
\end{center}}
\end{figure}
\begin{figure}[p]
{\begin{center}
 \includegraphics[width=0.9\textwidth]{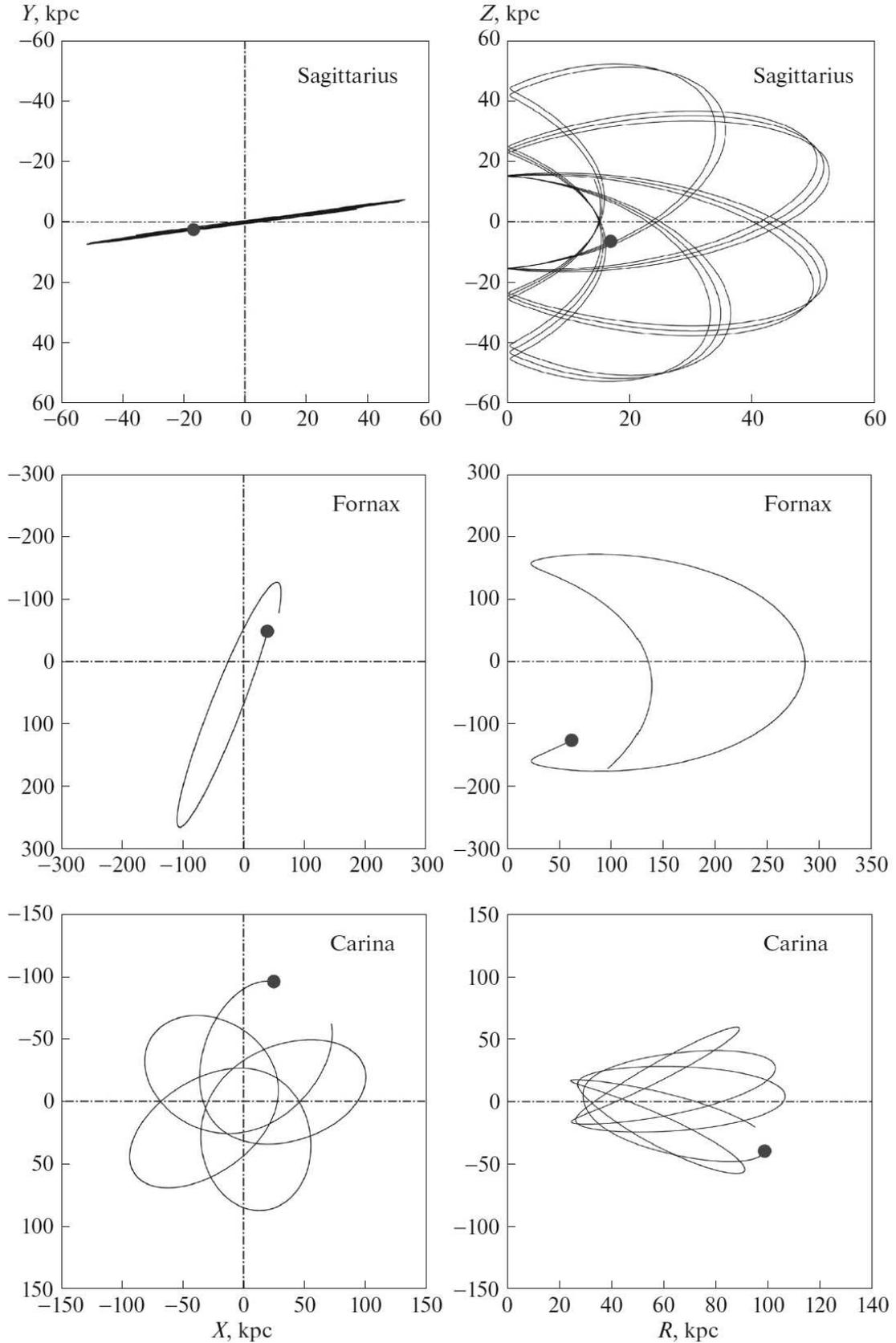}
 \caption{
Galactic orbits of the dwarf galaxies Sagittarius, Fornax, and
Carina over 10 billion years into the past.
  } \label{f456}
\end{center}}
\end{figure}
\begin{figure}[p]
{\begin{center}
 \includegraphics[width=0.9\textwidth]{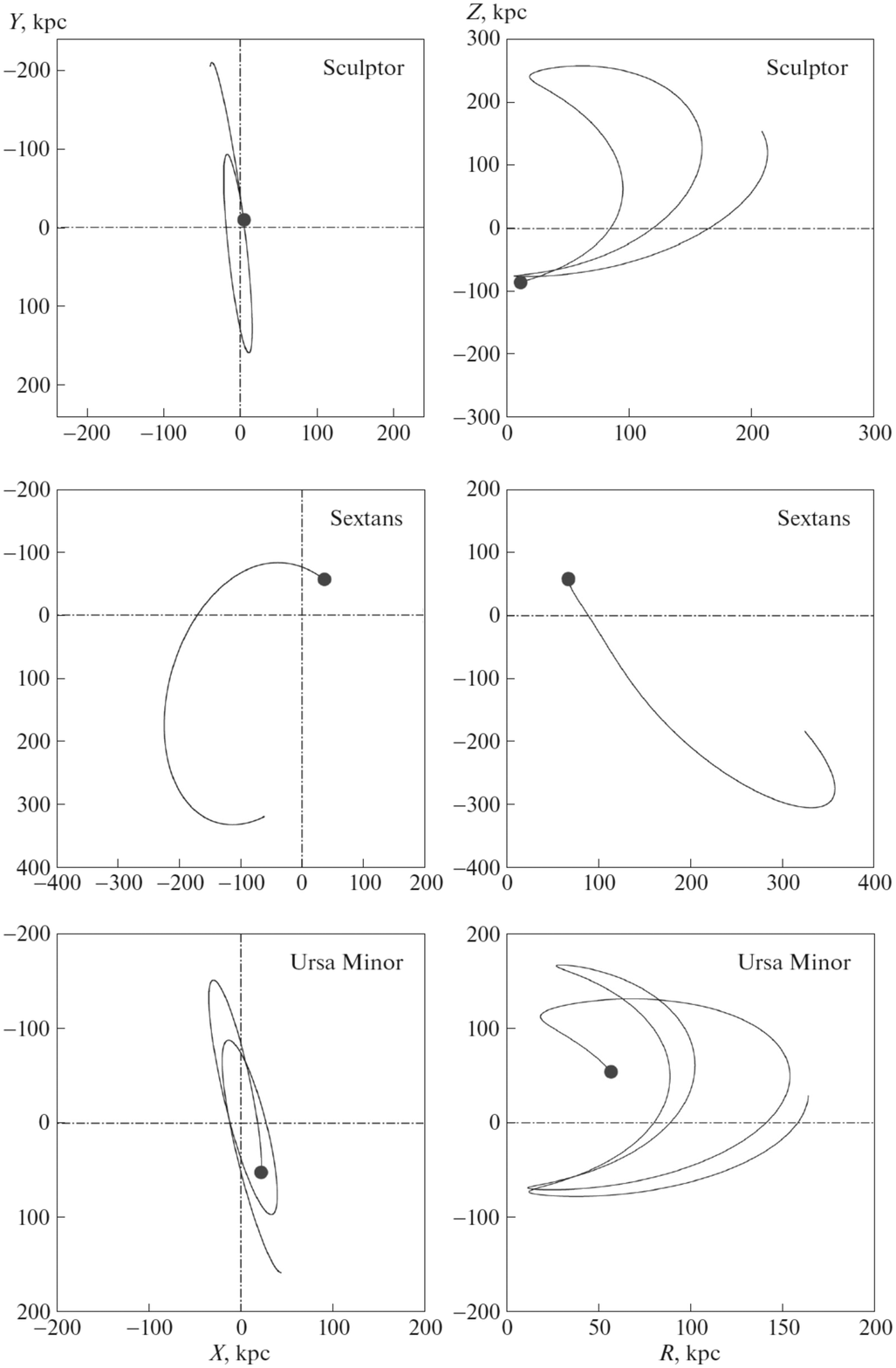}
 \caption{
Galactic orbits of the dwarf galaxies Sculptor, Sextans, and Ursa
Minor over 10 billion years into the past.
  } \label{f789}
\end{center}}
\end{figure}
\begin{figure}[p]
{\begin{center}
 \includegraphics[width=0.9\textwidth]{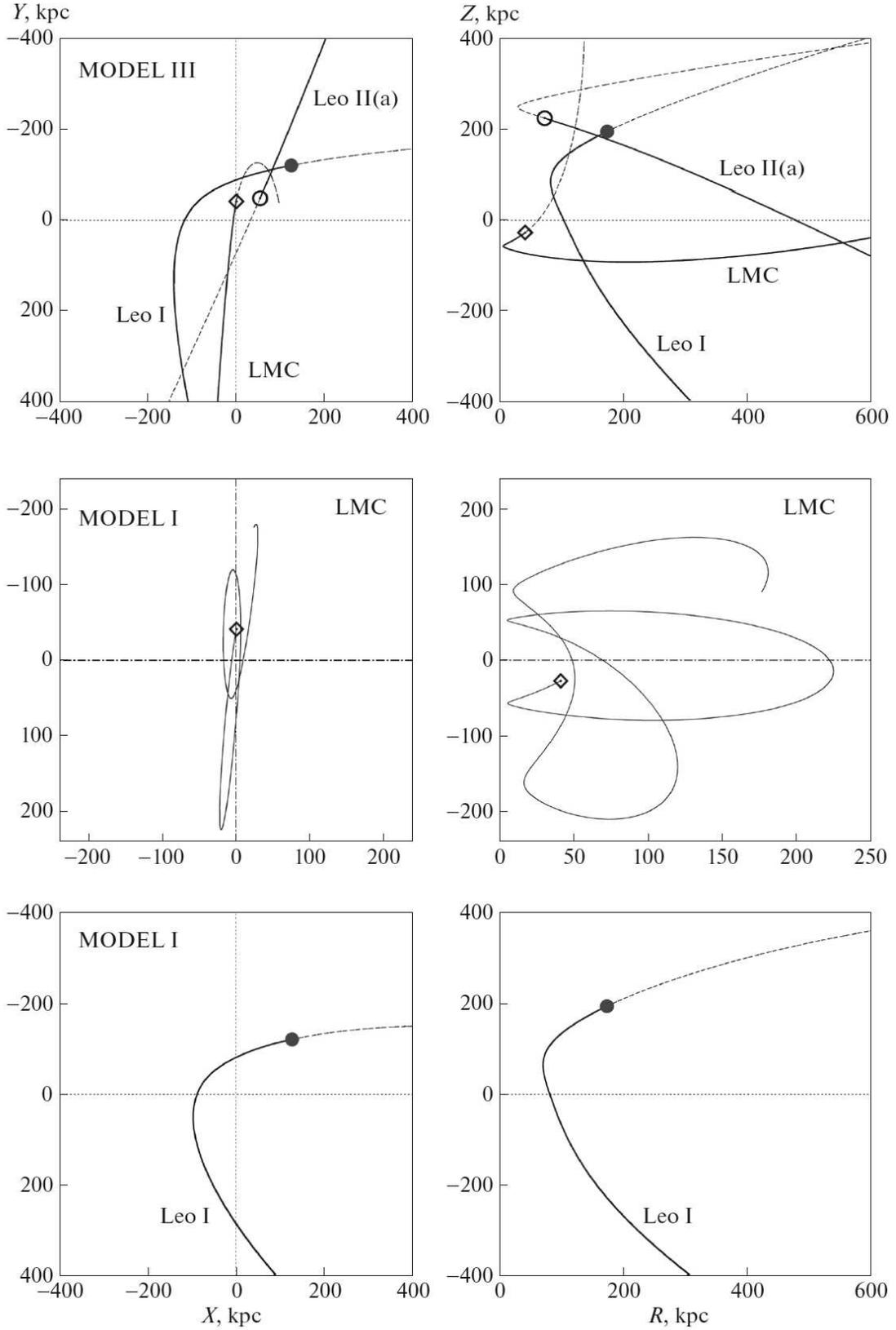}
 \caption{
Galactic orbits of the dwarf galaxies LMC, Leo I, and Leo II (a)
over 10 billion years into the past (solid curves) and 10 billion
years into the future (dashed curves). The trajectories in the
upper panel were computed using Model III for the Galactic
potential; the two other panels show results for Model I (with a
higher mass for the Galaxy).
  } \label{f4}
\end{center}}
\end{figure}
\begin{figure}[p]
{\begin{center}
 \includegraphics[width=0.9\textwidth]{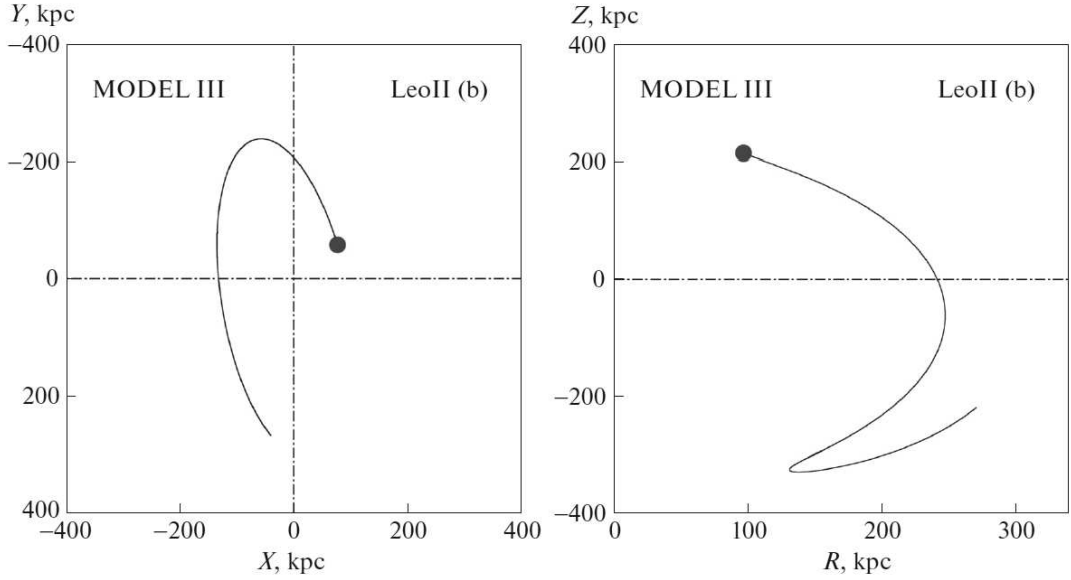}
 \caption{
Galactic orbits of the dwarf galaxy Leo II (b) over 10 billion
years into the past derived using Model III for the Galactic
potential.
  } \label{f5}
\end{center}}
\end{figure}
\begin{figure}[p]
{\begin{center}
 \includegraphics[width=0.9\textwidth]{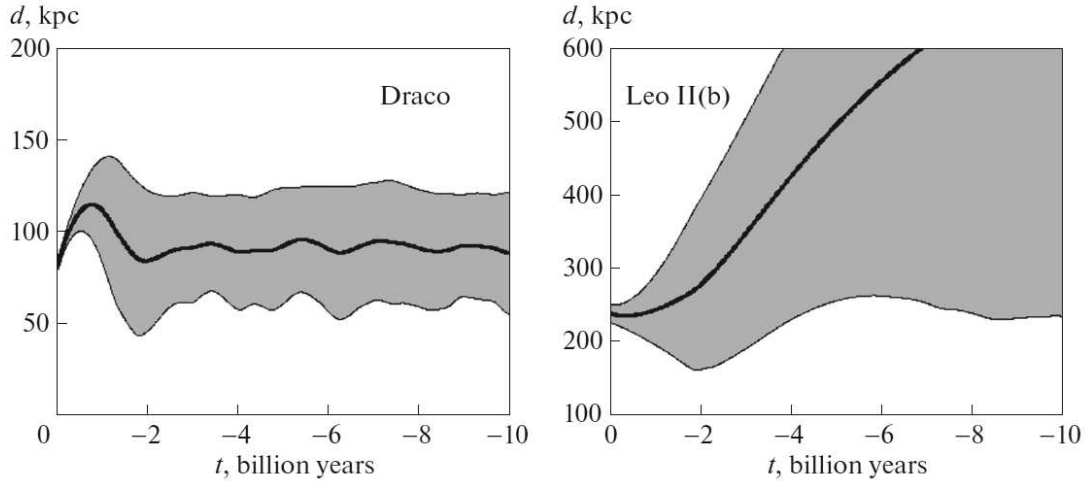}
 \caption{
Distances to Draco and Leo II (solid curves) as a function of
time, computed taking into account the uncertainties in the input
data and gravitational-potential parameters; the shading shows the
confidence area corresponding to the $\pm1\sigma$ uncertainties.
  } \label{f6}
\end{center}}
\end{figure}

 \subsection{RESULTS AND DISCUSSION}
The Galactic orbits of all 11 dwarf galaxies are displayed in
Figs. 1--4. We show two projections for each of the galaxies, onto
the $XY$ and $RZ$ planes.

We see from the upper part of Table 4 that the motion of eight of
the galaxies (Fornax, Sagittarius, Ursa Minor, Sculptor, Sextans,
Draco, Carina, and SMC) demonstrates that they are gravitationally
bound to the Milky Way. In particular, this is indicated by the
comparatively low eccentricities of their orbits $(e<0.8).$ This
conclusion is based on an analysis of the galactic orbits computed
using the gravitation potential with a comparatively low total
mass, $M_{200}=(0.75\pm0.19)\times10^{12}M_\odot.$

The upper part of Table 4 and Fig. 4 also show that the LMC, Leo
I, and Leo II (a) have highly elongated orbits with eccentricities
$e>0.8.$ In such cases, a potential with a higher mass for the
Galaxy is usually applied (e.g., [4]), especially since estimates
of the Galaxy’s total mass lie in a fairly wide range:
$M_{200}=(0.4-2.0)\times10^{12}M_\odot$~[17].

Some orbital characteristics of the Milky Way’s companion galaxies
were estimated in [5] using a model for the gravitation potential
with a total mass of the Galaxy $M_{200}=1.1\times10^{12}M_\odot$.
As in our study, high eccentricities, $e>0.8,$ were obtained for
the LMC, Leo I, and Leo II (a); see Fig. 8 in [5]. On the other
hand, the orbit derived for Leo II (b) with the new proper motions
has a low eccentricity, $e=0.23.$ The orbit of this galaxy derived
using Model III is shown in Fig. 5.

The lower part of Table 4 presents the results obtained by
applying Model I to the LMC and Leo I. Here, the total mass of the
Galaxy is $M_{200}=(1.45\pm0.30)\times10^{12}M_\odot$. This
yielded a much less elongated orbit for the LMC. However, although
the orbit of Leo I became smaller, it is still very elongated. Of
course, it is possible to address this problem in the opposite
sense, and determine the mass of the Galaxy for which Leo I would
be gravitationally bound to the Milky Way. A fairly high mass for
the Galaxy was obtained in [4] when considering a similar problem:
$M=2.4\times10^{12}M_\odot$ for the sample with Leo I, or
$M=1.7\times10^{12}M_\odot$ without Leo I.

Leo I has a very high radial velocity, indicating that it is only
weakly bound to the Galaxy. It is interesting that the orbital
characteristics for the LMC and Leo II (a) computed into the
future are virtually the same as those for the orbits computed
into the past. However, applying Model I to Leo I yields the
orbital parameters $a_{min}=261$ kpc, $a_{max}=1082$ kpc, and
$e=0.61.$ This leads us to expect that more accurate measurements
for Leo I could significantly influence the character of its
orbit, as occurred for Leo II.

The orbital parameters for all the dwarf-galaxy companions of the
Milky Way we have analyzed, obtained taking into account the
uncertainties in the input data and in the parameters of the
Galaxy’s gravitation potential, are presented in Table 5. As an
example of two typical cases of how the uncertainties influence
the orbits, Fig. 6 shows the distances to Draco and Leo II (b) and
their confidence areas corresponding to the $\pm1\sigma$ level of
uncertainties in the distances.

To determine the mean parameters and their rms deviations, we
carried out statistical Monte Carlo simulations using 100
independent realizations of the random errors for each object,
described by a normal law with zero mean and specified rms
deviation. The results (Table 5) can be used to judge the extent
to which a companion is gravitationally bound to the Galaxy,
taking into account the measurement uncertainties. For example,
the random errors have brought about a shift of the mean orbital
eccentricity towards higher values relative to the nominal value
for several of the galaxies (compare Tables 4 and 5); this is
especially clear for Sextans. For most of the galaxies (Fornax,
Sagittarius, Ursa Minor, Sculptor, Draco, Carina, the SMC), the
character of their orbits testifies that they are tightly
gravitationally bound to the Galaxy, even in the case of a
comparatively small total mass for the Galaxy,
$M_{200}=0.75\times10^{12}M_\odot$, corresponding to the modified
model for the gravitational potential of Navarro, Frenk, and White
[12].

 \subsection{CONCLUSIONS}
 We have compiled the results of high-accuracy
measurements of proper motions, radial velocities, and distances
for 11 galaxies in the Local Group from the literature. The most
distant of these, Leo I, is 254 kpc from the Sun.

We have derived Galactic orbits for all 11 galaxies covering a
long time interval. We mainly used the Navarro, Frenk, \& White
model for the Galactic potential, refined by us, with a
comparatively small total mass for the Galaxy,
$M_{200}=0.75\times10^{12}M_\odot.$ The character of the orbits of
most of these galaxies (Fornax, Sagittarius, Ursa Minor, Sculptor,
Draco, Carina, the SMC) demonstrates that they are tightly
gravitationally bound to the Galaxy.

However, the orbits for some of the galaxies were found to be
fairly elongated; for example, the orbital eccentricity derived
for the LMC is $e=0.90.$ Using another model for the Galactic
potential (the Allen-Santill\'an model, again refined by us), with
a total mass for the Galaxy $M_{200}=1.45\times10^{12}M_\odot$, we
obtained a more compact orbit for the LMC with the eccentricity
$e=0.65.$

An exception is the galaxy Leo I, whose orbit is extremely
elongated, $e>0.8,$ even using this latter model potential. Thus,
the question of whether this galaxy is fully gravitationally bound
to the Milky Way remains open.

 \subsubsection*{ACKNOWLEDGEMENTS}
We thank the referee for useful comments that have helped to
improve this paper. This work was supported by the Basic Research
Program P-7 of the Presidium of the Russian Academy of sciences,
under the subprogram ``Transitional and Explosive Processes in
Astrophysics''.

 \bigskip\subsubsection*{REFERENCES}

 {\small
\quad~ 1. A.W. McConnachie, Astron. J. 144, 4 (2012).

2. J. D. Diaz, S. E. Koposov, M. Irwin, V. Belokurov, and N. W.
Evans, Mon. Not. R. Astron. Soc. 443, 1688 (2014).

3. M. S. Pawlowski, S. S. McGaugh, and H. Jerjen, Mon. Not. R.
Astron. Soc. 453, 1047 (2015).

4. M. I. Wilkinson and N. W. Evans, Mon. Not. R. Astron. Soc. 310,
645 (1999).

5. M. Rocha, A. H. G. Peter, and J. Bullock, Mon. Not. R. Astron.
Soc. 425, 231 (2012).

6. H. Lux, J. I. Read, and G. Lake, Mon. Not. R. Astron. Soc. 406,
2312 (2010).

7. G. W. Angus, A. Diaferio, and P. Kroupa, Mon. Not. R. Astron.
Soc. 416, 1401 (2011).

8. C. Barber, E. Starkenburg, J. F. Navarro, A. W. McConnachie,
and A. Fattahi, Mon. Not. R. Astron. Soc. 437, 959 (2014).

9. S. L\'epine, A. Koch, R. M. Rich, and K. Kuijken, Astrophys. J.
741, 100 (2011).

10. M. S. Pawlowski and P. Kroupa, Mon. Not. R. Astron. Soc. 435,
2116 (2013).

11. R. P. van der Marel, Galaxy Masses as Constraints of Formation
Models, Ed. by M. Cappellari and S. Courteau (Cambridge University
Press, 2015), p. 1.

12. A. T. Bajkova and V. V. Bobylev, Astron. Lett. 42, 567 (2016).

13. C. Allen and A. Santill\'an, Rev. Mex. Astron. Astrofis. 22,
255 (1991).

14. M. I. Wilkinson and N. W. Evans, Mon. Not. R. Astron. Soc.
310, 645 (1999).

15. J. F. Navarro, C. S. Frenk, and S. D. M. White, Astrophys. J.
490, 493 (1997).

16. M. Miyamoto and R. Nagai, Publ. Astron. Soc. Jpn. 27, 533
(1975).

17. P. Bhattacharjee, S. Chaudhury, and S. Kundu, Astrophys. J.
785, 63 (2014).

18. C. Pryor, S. Piatek, and E. W. Olszewski, Astron. J. 149, 42
(2015).

19. D. I. Casetti-Dinescu and T. M. Girard, Mon. Not. R. Astron.
Soc. 461, 271 (2016).

20. K. Kinemuchi, H. C. Harris, H. A. Smith, N. A. Silbermann, L.
A. Snyder, A. P. LaCluyz\'e, and C. L. Clark, Astron. J. 136, 1921
(2008).

21. T. E. Armandroff, E. W. Olszewski, and C. Pryor, Astron. J.
110, 2131 (1995).

22. R. P. van der Marel and J. Sahlmann, Astrophys. J. 832, L23
(2016).

23. R. P. van der Marel and N. Kallivayalil, Astrophys. J. 781,
121 (2014).

24. A. G. A. Brown, A. Vallenari, T. Prusti, J. H. J. de Bruijne,
et al., Astron. Astrophys. 595, 2 (2016).

25. L. Lindegren, U. Lammers, U. Bastian, J. Hernandez, et al.,
Astron. Astrophys. 595, 4 (2016).

26. S. Piatek, C. Pryor, and E. W. Olszewski, Astron. J. 152, 166
(2016).

27. A. Koch, J. T. Kleyna, M. I. Wilkinson, E. K. Grebel, G. F.
Gilmore, N. Wyn Evans, R. F. G. Wyse, and D. R. Harbeck, Astron.
J. 34, 566 (2007).

28. M. Bellazzini, N. Gennari, and F. R. Ferraro, Mon. Not. R.
Astron. Soc. 360, 185 (2005).

29. A. Irrgang, B. Wilcox, E. Tucker, and L. Schiefelbein, Astron.
Astrophys. 549, 137 (2013).

30. R. Sch\"onrich, J. Binney, and W. Dehnen, Mon. Not. R. Astron.
Soc. 403, 1829 (2010).
 }
 \end{document}